\documentclass[aps,showpacs,twocolumn,longbibliography,nofootinbib,secnumarabic,superscriptaddress]{revtex4-1}

\usepackage{graphicx}
\usepackage{dcolumn}
\usepackage{bm}
\usepackage{braket}
\usepackage{leftidx}
\usepackage{cancel}
\usepackage{amsmath}
\usepackage{epstopdf}
\usepackage{color}
\usepackage[mathcal]{eucal}
\usepackage{float}

\usepackage{yfonts}
\usepackage{color}

\usepackage{scalerel}
\usepackage{mathtools}

\begin{document}
\date{\today}

\title{Fano Enhancement of Unlocalized Nonlinear Optical Processes} 

\author{Mehmet G\"{u}nay}
\affiliation{Faculty of Arts and Science, Burdur Mehmet Akif Ersoy University, 15030 Burdur, Turkey}
\affiliation{Institute of Nuclear Sciences, Hacettepe University, 06800, Ankara, Turkey}

\author{Ahmet Cicek}
\affiliation{Faculty of Arts and Science, Burdur Mehmet Akif Ersoy University, 15030 Burdur, Turkey}

\author{Nurettin Korozlu}
\affiliation{Faculty of Arts and Science, Burdur Mehmet Akif Ersoy University, 15030 Burdur, Turkey}

\author{Alpan Bek}
\affiliation{Department of Physics, Middle East Technical University, 06800 Ankara, Turkey}
\affiliation{Center for Solar Energy Research and Applications, Middle East Technical University, 06800 Ankara, Turkey}

\author{Mehmet Emre Tasgin}
\affiliation{Institute of Nuclear Sciences, Hacettepe University, 06800, Ankara, Turkey}

\date{\today}

\begin{abstract}
Field localization boosts  nonlinear optical processes at the hot spots of metal nanostructures. Fano resonances can further enhance these ``local" processes taking place at the hot spots. However, in the conventional nonlinear materials, the frequency conversion takes place along the entire crystal body. That is, the conversion process is ``unlocalized''. The path interference (Fano resonance) schemes developed for localized processes become useless in such materials. Here, we develop Fano enhancement schemes for unlocalized nonlinear optical processes. We show that a 3 orders of magnitude Fano enhancement {\it multiplies} the enhancements achieved via field trapping techniques, e.g., in epsilon-near-zero~(ENZ) materials. We demonstrate the phenomenon both analytically and by numerical solutions of Maxwell's equations. The match between the two solutions is impressive. We observe that the interference scheme for unlocalized processes is richer than the one for the local processes. The method can be employed to any kind of nonlinear optical conversion. Moreover, the Fano enhancement can be continuously controlled by an applied voltage.
\end{abstract}

\maketitle

\section{Introduction}

Past two decades witnessed fascinating progresses in two research areas, classical and quantum plasmonics~\cite{tame2013quantum}. A phenomenon is highlighted commonly in both research areas: strong field localization at the metal nanoparticle~(MNP) hot spots.  The localized field intensity can be 5-7 orders of magnitude stronger than the intensity of the incident light in the nm-size hot spots~\cite{stockman2011nanoplasmonics,hoppener2012self}. The (classical) plasmonics employs the field enhancement for ultra-high sensing~\cite{anker2008biosensing}, hot spot-size optical resolution~\cite{SNOM_Nature_2003,zhang2013chemicalmapping}, high-frequency communication technologies~\cite{giannini2011plasmonic}, and for achieving extreme nonlinearity enhancements~\cite{kauranen2012nonlinear}. Nonlinear optical processes taking place at the hot spots can be enhanced with the second power of the local intensity, because both the input and the converted fields are localized~\cite{Double_resonance_SHG_PRB2017,ye2012plasmonic,chu2010double}. This makes plasmonics an attractive field for nonlinear optics.

Quantum plasmonics utilizes the strong field localization for enhanced light-matter interaction. When a quantum emitter~(QE) ---e.g., a molecule, a quantum dot~(QD) or a color-center--- is placed at the hot spot, it creates path interference effects called Fano resonances~\cite{pelton2019strong,liu2009plasmonic,leng2018strong,wu2010quantum,PlasIndTransPRL2008}. Fano resonances can provide control both over linear~\cite{LavrinenkoPRB2019} and nonlinear~\cite{butet2014fano,thyagarajan2013augmenting,butet2012nonlinear,turkpence2014engineering,paspalakis2014strongly,singh2016enhancement}  response of metal nanostructures. The nonlinearity enhancement provided by the local field enhancement~(in classical plasmonics) can further be multiplied by a Fano-enhancement factor which can be 3 orders of magnitude depending on the choice of the QE level spacing.



Fano resonances, studied so far, control the ``local'' nonlinear processes that take place at the hot spots. However, conventional frequency converters are nonlinear crystals where conversion takes place along the entire crystal body. The path interference (Fano) enhancement schemes for localized nonlinear processes become useless for such nonlinear crystals. In this work, we theoretically and computationally demonstrate how a Fano-control of such ``unlocalized'' nonlinear processes can be made possible.

Nonlinear processes taking place in conventional frequency converters can also be enhanced by concentrating the field into the crystals. Embedding plasmonic nanoparticles~(NPs) into nonlinear crystals can enhance the conversion~\cite{nie2018plasmonic,li2017giant} by localizing the field near the NPs.  A better method employs epsilon-near-zero~(ENZ) materials. An ENZ medium not only greatly enhances the field inside the frequency converter~\cite{ENZnonlinearNatureReview2019,capretti2015comparative,luk2015enhanced,deng2020giant}, but also relieves the phase-matching condition~\cite{kauranen2013phasematching} in longer nonlinear crystals. Moreover, recently explored longitudinal ENZ (LENZ) materials can provide stronger second harmonic~(SH) and third harmonic~(TH) conversion enhancements via circumventing material losses~\cite{vincentiLENZ2017}.  In these (more convenient) materials, enhanced field is confined in the body of the material rather than being localized at nm-size hot spots. One can also combine the two field trapping methods~\cite{ahmadivand2018toroidal} for stronger enhancement rates. These methods rely on the enhancement of the field inside the nonlinear crystal. Presence of an unlocalized Fano-enhancement method, compatible with such materials and multiplying all other enhancement factors, would be very beneficial in achieving high conversion factors in nonlinear crystals.

In this paper, we develop such a Fano-control mechanism for the \textit{further} enhancement of ``unlocalized'' nonlinear optical processes taking place in conventional frequency converters~\cite{nikogosyan2006nonlinear}. We show that MNP-QE dimers can introduce path interference~(Fano) effects in unlocalized nonlinear processes. 
With appropriate choice of the QE's level spacing~$\omega_{\rm \scriptscriptstyle{QE}}$, nonlinear process can be enhanced, e.g., by 3 orders of magnitude. This enhancement takes place {\it without} increasing the fundamental (first harmonic) field intensity in the crystal. In other words, our method further enhances ($\sim$1000$\times$) the nonlinear field which is already enhanced via field trapping techniques, e.g., in MNP-doped~\cite{nie2018plasmonic,li2017giant} and ENZ/LENZ materials~\cite{ENZnonlinearNatureReview2019,capretti2015comparative,deng2020giant,vincentiLENZ2017}. More explicitly, for instance, $10^4$ times conversion enhancement in an ENZ material~\cite{ENZnonlinearNatureReview2019,capretti2015comparative,deng2020giant,vincentiLENZ2017} can be further multiplied by a factor of $\sim$1000, yielding a total enhancement ratio of $10^7$. The method can be applied to  both photonic and nanophotonic (smaller) devices.

We consider a system where MNP-QE dimers are embedded in a crystal body, see Fig.~\ref{fig1}a. We utilize the MNPs as strong interaction centers for the ``converted'' (e.g., SH) field. MNPs collect the \textit{unlocalized} converted field into the hot spots and make it interact with the QE. Intriguingly, we find that MNP-QE dimers can control the produced nonlinear field throughout the entire crystal body.

Our numerical (COMSOL~\cite{comsol}) simulations clearly demonstrate that the SH ($2\omega$) field intensity can be enhanced by a factor of $\sim$1000 {\it without increasing} the fundamental frequency~(FF, $\omega$) field inside the crystal. We not only demonstrate the phenomenon via numerical solutions of the Maxwell's equations, but also explain the physics behind the Fano enhancement mechanism which multiplies the enhancements created by the field trapping techniques~\cite{nie2018plasmonic,li2017giant,ENZnonlinearNatureReview2019,capretti2015comparative,luk2015enhanced,deng2020giant,vincentiLENZ2017,ahmadivand2018toroidal}. We present a simple analytical expression for the converted (SH) field amplitude. We demonstrate that further enhancement occurs due to the cancellations in the denominator of this expression without changing the FF field intensity. The results for the numerical solutions of the Maxwell's equations and the analytical model match successfully. The analytical model also shows that the interference scheme for the control of the ``unlocalized'' nonlinear processes is different than the scheme for the control of localized processes. It provides a richer cancellation scheme.

It is worth noting that we refer to the ``Fano enhancement'' as the further enhancement which multiplies other enhancement factors appearing due to the field trapping techniques. As explicitly demonstrated in the text, Fano enhancement appears due to the cancellations in the denominator of the nonlinear response and is independent from the FF field enhancements. The total enhancement includes both effects.

 The exceptional features of our method can be summarized as follows. (i) The path interference (Fano) can further enhance an unlocalized nonlinear process which is already enhanced via field trapping techniques~\cite{nie2018plasmonic,li2017giant,ENZnonlinearNatureReview2019,capretti2015comparative,luk2015enhanced,deng2020giant,vincentiLENZ2017,ahmadivand2018toroidal}. (ii) Fano enhancement scheme does not increase the FF~($\omega$) field intensity. This allows a $\sim$1000 times further enhancement for the crystals already operating in the upper temperature limits. (iii) Since a ~$\sqrt{1000}\simeq$33 times smaller pump intensity would suffice for efficient frequency conversion, battery lifetimes of portable lasers would be extended significantly.  (iv) Moreover, the Fano enhancement factor can be \textit{continuously tuned} around its peak value via an applied voltage. The applied voltage could tune the QE resonance $\omega_{\rm \scriptscriptstyle{QE}}$ which in turn could be used to switch the Fano enhancement. (v) Above all, the Fano-enhancement scheme we demonstrate for the SHG process can be employed also for higher order nonlinear optical processes. For instance,  we demonstrate the path interference schemes also for third harmonic generation and four-wave mixing in the Supplementary Materials.

Here, we work out the path interference effects in the nonlinear response of materials. A linear version of the effect would be associated with the Fano lasers~\cite{MorkPRL2014FanoLaser,yu2017demonstration} if one removes the MNP from the interference scheme. A photonic crystal Fano laser made of a line defect waveguide (active medium) coupled to a narrow linewidth nanocavity  is studied in literature. The nanocavity introduces the Fano effect which enables an interesting self-pulsing phenomenon~\cite{yu2017demonstration} and a high frequency modulation~\cite{MorkPRL2014FanoLaser}. In such a photonic crystal system, one can also study the nonlinear Fano effect by choosing the nanocavity resonance near the converted frequency. In such a system, however, the nanocavity resonance is not voltage-tunable.  Thus, the nanocavity resonance has to be manufactured carefully and it is hard to arrange the Fano resonance peak.


\section{Results}

\subsection{Dynamics of the system}

\begin{figure}
\centering
\includegraphics[width=0.47\textwidth]{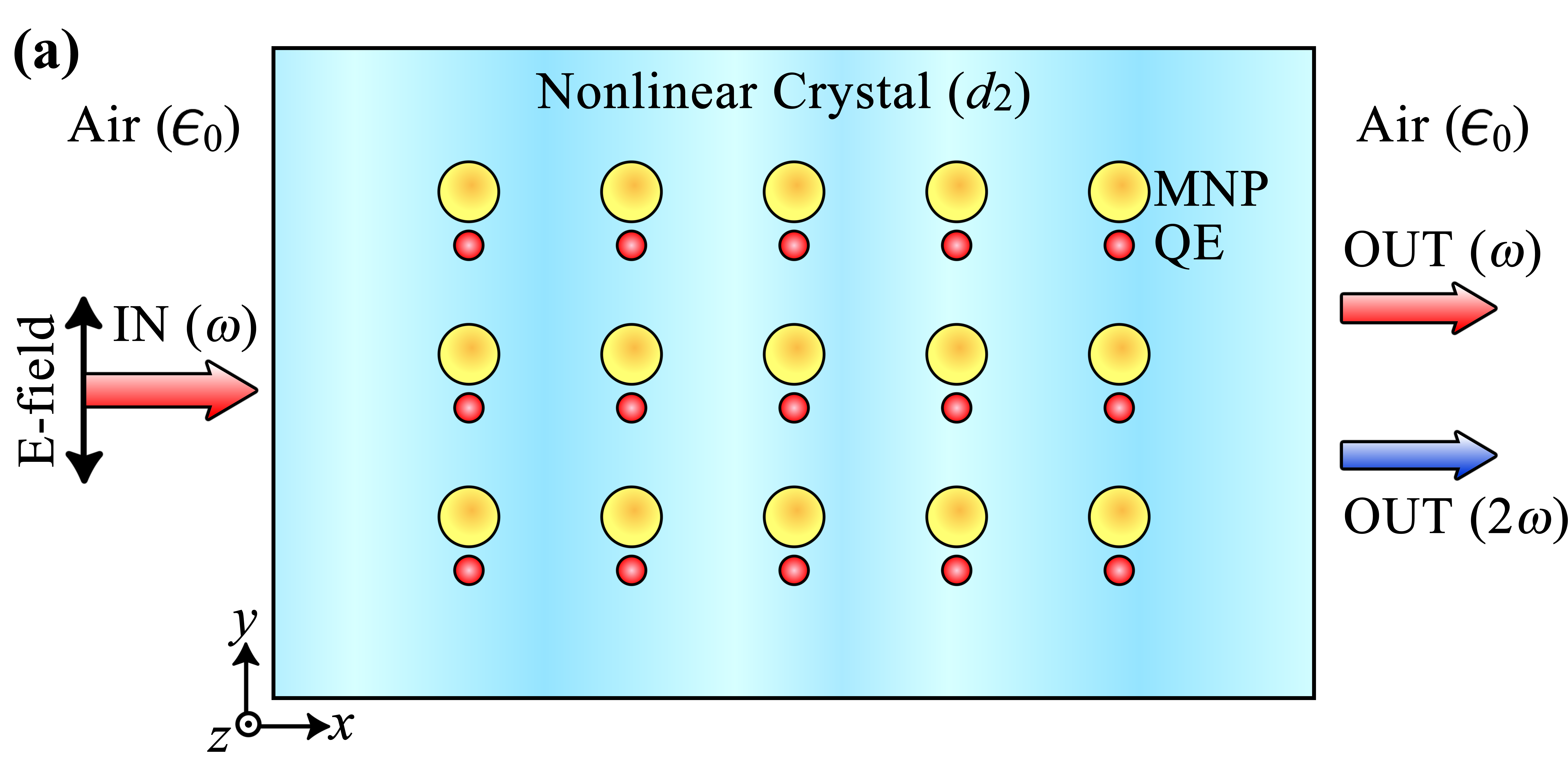} 
\hspace{1 cm}
\includegraphics[width=0.43\textwidth]{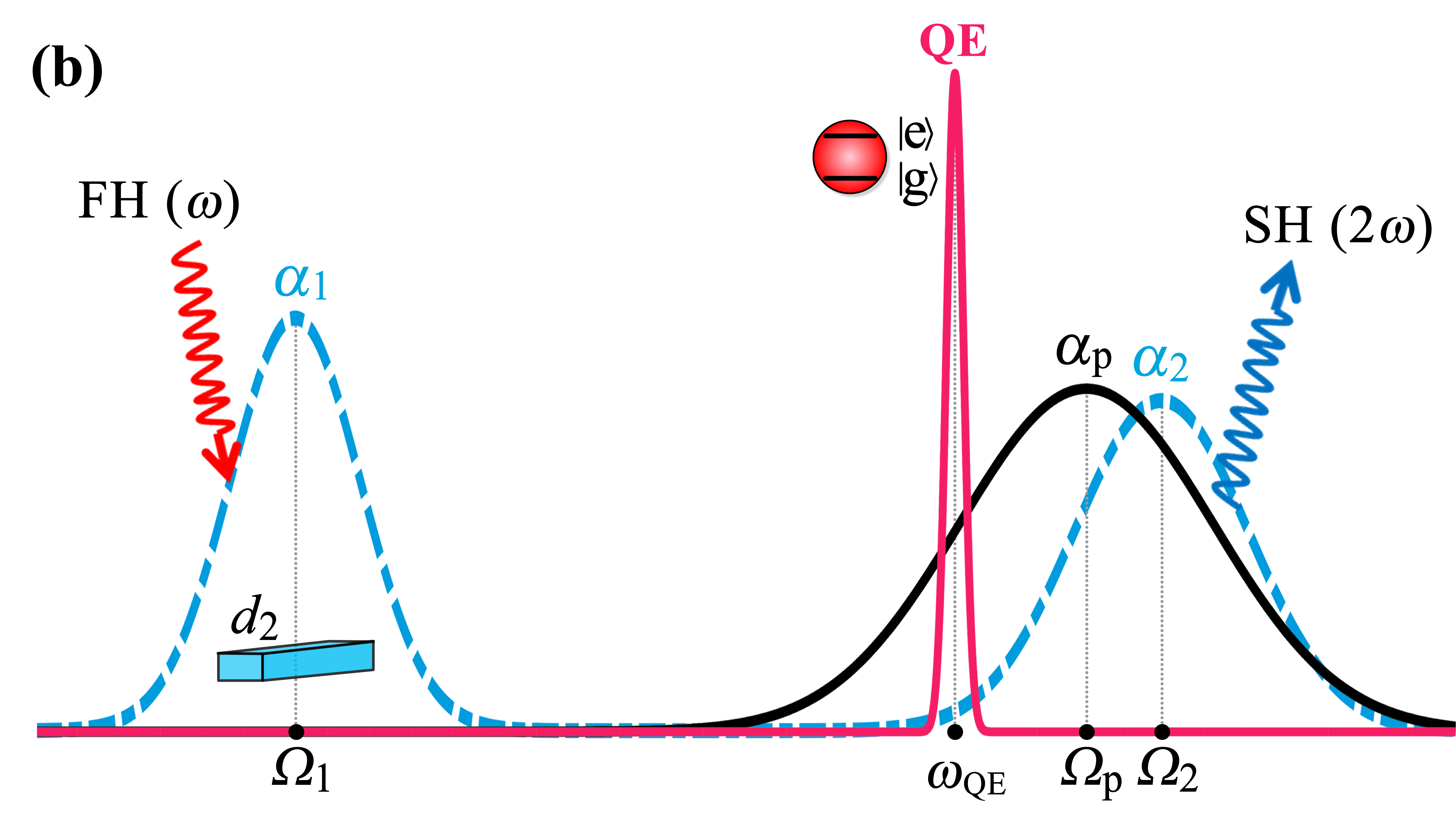} 
\caption{{\bf Fano enhancement scheme for a second harmonic converting crystal.} (a) MNP-QE dimers are embedded in a nonlinear crystal with a second order susceptibility $d_2$. The dimers interact with the converted (2$\omega$) field and introduce path interference (Fano) effects. The crystal is pumped with a y-polarized laser of frequency $\omega$. The nonlinear process can be enhanced without exploiting field trapping techniques. (b) The diagram for the crystal~($\alpha_{1,2}$) and the plasmon~($\alpha_p$) modes of resonances $\Omega_{1,2}$ and $\Omega_p$, and the QE resonance $\omega_{\rm \scriptscriptstyle QE}$. The converted field interacts with the MNP~($\alpha_p$) and the MNP couples with the QE.}
\label{fig1}
\end{figure}

In particular, we consider the SHG process from a nonlinear crystal, see Fig.~\ref{fig1}a, in which MNP-QE dimers are embedded for the Fano enhancement. The dynamics of the system and the interactions can be described as follows.

A laser of frequency $\omega$ (fundamental, $\lambda=$ 1064 nm) pumps the $\alpha_1$-mode of the nonlinear crystal. The pump excites the $\omega$ frequency photons in the $\alpha_1$-mode, see Fig.~\ref{fig1}b. The resonance of the $\alpha_1$ crystal mode is $\Omega_1$. Here, $\alpha_1$ refers to the amplitude of the fundamental frequency~(FF) field inside the crystal. The crystal performs SHG: two $\omega$ photons in the crystal ($\alpha_1$-mode) combine to generate a $2\omega$ photon ($\lambda/2=$ 532 nm) in the $\alpha_2$-mode of the crystal. The resonance of the $\alpha_2$ crystal mode is $\Omega_2$. $\alpha_2$ refers to the amplitude of the SH field inside the crystal. There can exist other modes in the crystal, but we refer merely to the relevant ones: the pumped mode $\alpha_1$ and the mode $\alpha_2$ into which SHG takes place. 

The resonance frequency of the MNP plasmon mode $\Omega_p$ is chosen about the SH frequency $2\omega$, so $\Omega_p$ is also about the resonance frequency of the $\alpha_2$ crystal mode $\Omega_2$, see Fig.~\ref{fig1}b. The SH generated field interacts strongly (of strength $g$) with the MNP. The MNP localizes the generated SH field into its hot spot as a plasmonic near-field. $\alpha_p$ refers to the amplitude of the plasmon excitation on the MNP. The plasmon mode displays a localized near-field at the MNP hot spot. The QE is placed at this hot spot. Thus, the QE interacts strongly (of strength $f$) with the plasmon mode in the near-field of the MNP. This way, the MNP-QE dimer creates a Fano resonance effect on the SHG process which takes place along the entire crystal body.

\subsection{Analytical Model}

We first study the coupled system of the nonlinear crystal and the MNP-QE dimer with a basic analytical model. We obtain a simple analytical expression for the second harmonic~(SH) field amplitude $\alpha_2$.

The equations of motion~(Eqs.~(\ref{EOMa})-(\ref{EOMe}) in the Appendix) governing the dynamics of the system can be obtained from the hamiltonian [Eq.~(\ref{Htot})] describing the system. One also includes the decay rates for the fields and the QE into the equations of motion. Derivations can be found in the Appendix.

In the steady state, the SH field amplitude $\alpha_2$ can be obtained as
\begin{equation}
{\alpha}_2=\frac{-i \chi_2}{[i(\Omega_2-2\omega)+\gamma_2 ] +\frac{|g|^2}{[i(\Omega_p-2\omega)+\gamma_p ] - \frac{|f|^2y}{i(\omega_{\rm \scalebox{0.4}{QE}}-2\omega)+\gamma_{{}_{\rm \scalebox{0.4}{QE}}}}}} {\alpha}_1^2.
\label{alpha2}
\end{equation}
Here, $|\alpha_2|^2$ gives the number of SH generated $2\omega$ photons in the crystal. Similarly, $|\alpha_1|^2$ is the number of fundamental frequency~(FF, $\omega$) photons in the crystal. $\chi_2$ is a constant (an overlap integral~\cite{gunay2020controlling}) proportional to the second-order susceptibility of the nonlinear crystal, i.e., $\chi_2 \propto d_2$. $y=\rho_{ee}-\rho_{gg}$ is the population inversion of the QE, where $\rho_{ee}$ and $\rho_{gg}$ are the probabilities for the QE to be in the excited and the ground state, respectively, with the constraint $\rho_{ee}+\rho_{gg}=1$. $\gamma_{1,2}$, $\gamma_p$ and $\gamma_{\scriptscriptstyle{\rm QE}}$ are the decay rates for of the two crystal modes, the plasmon mode and the QE, respectively. $\Omega_{1,2}$, $\Omega_{p}$ and $\omega_{\scriptscriptstyle{\rm QE}}$ are the resonances in the same order.

\paragraph*{Fano enhancement}

Eq.~(\ref{alpha2}) reveals a striking mechanism for the Fano enhancement. Namely, one can increase the number of SH generated photons $|\alpha_2|^2$ ``without'' increasing the number of photons in the FF $\omega$, $|\alpha_1|^2$. This can simply be performed by introducing cancellations in the denominator of Eq.~(\ref{alpha2}).

On the one hand, $|\alpha_2|^2$ can be enhanced via enhancing the first harmonic field $|\alpha_1|^2$ using field trapping techniques~\cite{nie2018plasmonic,li2017giant,ENZnonlinearNatureReview2019,capretti2015comparative,luk2015enhanced,deng2020giant,vincentiLENZ2017}. One increases $|\alpha_1|^2$, for instance, by using an ENZ material. On the other hand, the enhancement due to the cancellations in the denominator of Eq.~(\ref{alpha2}) ``multiplies'' the enhancement of the $|\alpha_1|^2$. The latter enhancement (owing to the denominator) is called as the Fano enhancement. In advance, we state that in our numerical simulations using the analytical model, $|\alpha_1|^2$ does not change, thus, the enhancement in $|\alpha_2|^2$ originates solely from the denominator of Eq.~(\ref{alpha2}). The number of photons in the $\alpha_2$-mode is always (substantially) smaller than the number of photons of the FF $\alpha_1$-mode.

The cancellation scheme works as follows. The first term in the denominator, $i(\Omega_2-2\omega)+\gamma_2$, belongs to the bare crystal. That is, only this term exists if the nonlinear crystal is not embedded with a MNP-QE dimer. The second term in the denominator, $|g|^2/(\ldots)$, appears due to the MNP-QE dimer. If one arranges $g$, $f$, $\omega_{\scriptscriptstyle{\rm QE}}$ appropriately, the $|g|^2/(\ldots)$ term is able to cancel the first term partially. This can reduce the denominator in Eq.~(\ref{alpha2}), thus enhance the production of the SH 2$\omega$ photons $|\alpha_2|^2$ without relying to an enhancement in $|\alpha_1|^2$.
  
\begin{figure}
\includegraphics[width=0.47\textwidth,trim=0cm 0.5cm 0.0cm 0.5cm,clip]{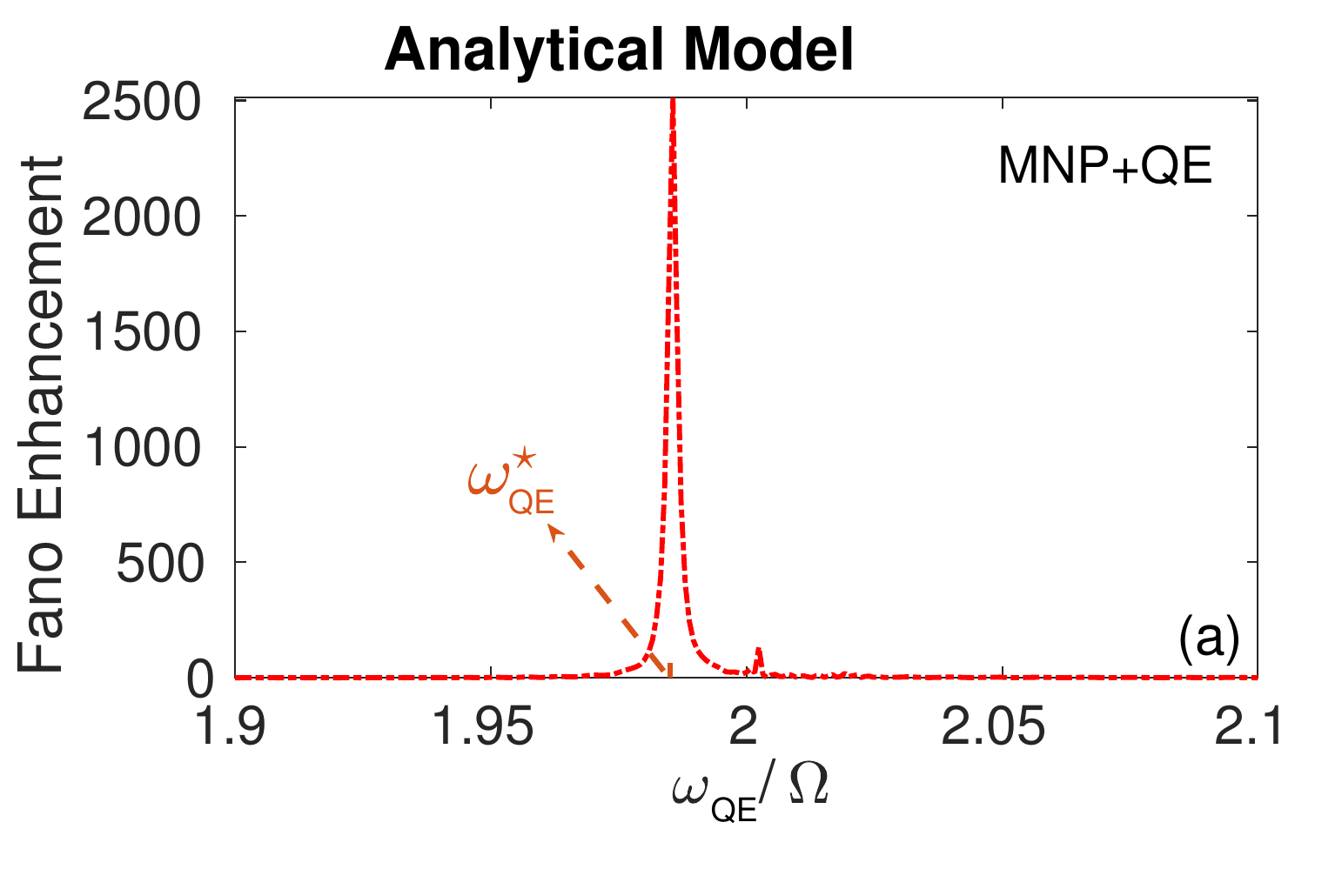} 
\includegraphics[width=0.47\textwidth,trim=0cm 1cm 0.0cm 1.5cm,clip]{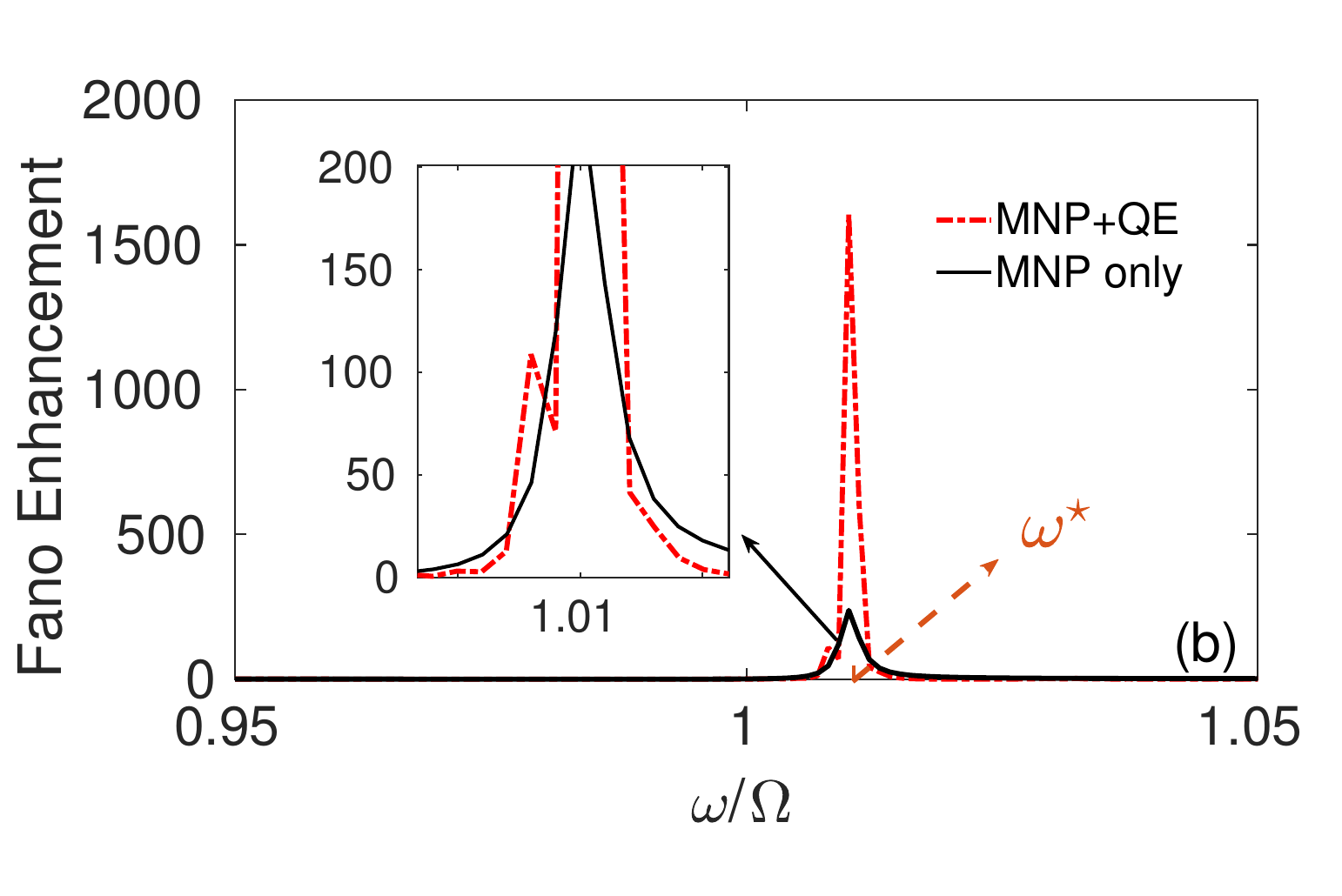}
\caption{{\bf Fano enhancement factors predicted by the  analytical model.} (a) We scan different values of the QE level spacing $\omega_{\scriptscriptstyle{\rm QE}}$ in order to find the particular value $\omega_{\scriptscriptstyle{\rm QE}}^{\star}\simeq$1.98$\Omega$ which performs cancellations in the denominator of $\alpha_2$ in Eq.~(\ref{alpha2}). Fano enhancement factor reaches $~\sim$2500. (b) We also fix the QE at $\omega_{\scriptscriptstyle{\rm QE}}$=2$\Omega$ and scan the pump frequency $\omega$ in order to find the particular value $\omega^{\star}$=1.01$\Omega$ which performs cancellations in the denominator of $\alpha_2$ in Eq.~(\ref{alpha2}). MNP-QE dimer (red dashed line) can achieve a Fano enhancement factor of $\sim$1500. Using only MNPs~(black line) can also introduce a Fano enhancement factor of $\sim$200. The FF field $|\alpha_1|^2$ do not change (not depicted). The frequencies are scaled with the frequency of the $\Lambda$=1064 nm light, i.e., $\Omega=2\pi c/$1064 nm.  
}
\label{fig2}
\end{figure}

In Fig.~\ref{fig2}, we present the enhancement which takes place merely due to the Fano resonance. We keep the laser (pump) strength constant and record the enhancement in $|\alpha_2|^2$. We observe that the SH intensity $|\alpha_2|^2$ can enhance $\sim$1000 times while the FF intensity $|\alpha_1|^2$ does not change. We calculate the amplitudes $\alpha_{1,2}$ via time evolution of the equations of motion for the system, see Eqs.~(\ref{EOMa})-(\ref{EOMe}) in the Appendix.


Throughout the paper, we scale frequencies by the constant $\Omega$, the frequency of the 1064 nm light, i.e., $\Omega=2\pi c/\Lambda$ with $\Lambda$=1064 nm. In Fig.~\ref{fig2}a, we set the laser (pump) frequency to $\omega=\Omega$ (1064 nm) and vary the level spacing of the QE, $\omega_{\scriptscriptstyle{\rm QE}}$. We observe that for the choice of $\omega_{\scriptscriptstyle{\rm QE}}$=$\omega_{\scriptscriptstyle{\rm QE}}^{\star}\simeq$1.98$\Omega$, the number of SH photons $|\alpha_2|^2$ is Fano enhanced $\sim$2000 times via cancellations in the denominator of Eq.~(\ref{alpha2}). (We use the star symbol ${\star}$ for indicating a value where maximum/optimum enhancement is achieved.)  $|\alpha_1|^2$ does not change (not depicted). In Fig.~\ref{fig2}b, we fix the QE level spacing to $\omega_{\scriptscriptstyle{\rm QE}}$=2$\Omega$ and  this time vary the frequency of the pump laser $\omega$. We observe that for the pump frequency $\omega$=$\omega^{\star}\simeq$1.01$\Omega$, the SHG $|\alpha_2|^2$ is Fano enhanced $\sim$1500 times without increasing the FF field intensity $|\alpha_1|^2$.

In Fig.~\ref{fig2}, we use the parameter set $\Omega_p=1.77\Omega$, $\gamma_p=0.1\Omega$, $g=f=0.1\Omega$, $\Omega_1=1.007\Omega$, $\Omega_2=2.013\Omega$, $\gamma_{1,2}=0.005\Omega$ and $\chi_2=10^{-4}\Omega$ (e.g., a small converter). The particular value of $\chi_2$ does not change the enhancement factors in our simulations.  In Fig.~\ref{fig2}a we set $\omega=\Omega$. In Fig.~\ref{fig2}b we set $\omega_{\rm \scriptscriptstyle QE}=2\Omega$.   These parameters are chosen according to the numerical simulations we conduct in the following section. For instance, the resonance frequency of the MNP $\Omega_p$ is set to $\lambda_p$= 600 nm ($\Omega_p=2\pi c/\lambda_p$). This is the resonance frequency of the gold NP plasmon calculated in an $n_{\scriptscriptstyle{\rm KDP}}$=1.51 index medium using the experimental dielectric function of gold, see the Supplementary Materials. The decay rates are chosen close to the ones for the crystal, plasmon and QE linewidths. We keep the parameters close to the ones for the numerical simulations, because we aim a {\it basic} comparison between the analytical and numerical (Maxwell) solutions. That is, we aim to check if the Fano enhancement appears using similar parameters in the analytical and the numerical simulations. In the next section, we show that an excellent match occurs between the two results.


 We can state the usefulness of the analytical model as follows. In a numerical (Maxwell) simulation alone, it is not possible to differentiate between a Fano enhancement and an enhancement due to local field improvement. Our analytical treatment sheds light onto the numerical results. This is because, analytical model does not take the localization effects (e.g., the change in the density of states near a MNP) into account. Thus, the enhancement factors presented by the analytical model are the ones only due to the Fano enhancement. This provides a useful tool in understanding the origins of the SHG enhancement. One can appreciate this better in the treatment below, where a MNP is shown to give rise to Fano enhancement without a QE. We also demonstrate this phenomenon using numerical simulations.
 
The presented analytical model is also valid (even to a higher extent) when the MNP-QE dimers are placed on the surface of the nonlinear crystal. In this case, MNP couples with the evanescent waves of the $\alpha_2$ mode. Strong coupling between the evanescent waves and the metal nanostuctures decorated on the crystals is a well reported phenomenon~\cite{fevrier2012giant,wang2017modulation,abdulhalim2018coupling,rashed2020hot}.  

We further observe that the interference scheme (i.e., the cancellations in the denominator) for the Fano control of an unlocalized system is different than the one for the localized nonlinear processes. In the Fano control of a local process (when the nonlinear process takes place at the hot spot), the denominator contains only the ${[i(\Omega_2-2\omega)+\gamma_2 ] - \frac{|f|^2y}{i(\omega_{\scalebox{0.4}{QE}}-2\omega)+\gamma_{\scalebox{0.4}{QE}}}}$ term~\cite{turkpence2014engineering}. In Eq.~(\ref{alpha2}), we observe that the cancellation scheme is richer.

\paragraph*{Fano enhancement using only MNPs}

A MNP is a strongly absorbing material with a broad linewidth. However, the analytical model below shows that a MNP alone can also introduce Fano enhancement effects. In order to study this phenomenon, we simply set the MNP-QE coupling to zero in Eq.~(\ref{alpha2}), i.e., $f$=0, and obtain the SH amplitude
\begin{equation}
{\alpha}_2=\frac{-i \chi_2}{[i(\Omega_2-2\omega)+\gamma_2 ] +\frac{|g|^2}{[i(\Omega_p-2\omega)+\gamma_p ]}} {\alpha}_1^2.
\label{alpha2_MNP}
\end{equation}
Here, $|g|^2/(\ldots)$ term can perform cancellations in the (first) term belonging to the bare crystal. The cancellation can be performed for the proper choices of the $\Omega_p$, $|g|^2$ and $\omega$. In Fig.~\ref{fig2}b, we demonstrate the phenomenon by varying the pump frequency $\omega$ for a fixed $\Omega_p=1.773\Omega$ (600 nm) and $g=0.1\Omega$. The black line shows a 200 Fano enhancement factor. In Fig.~3, the pump frequency is fixed at $\omega=\Omega$ and the MNP resonance exhibits a $\sim$200 times Fano enhancement at $\Omega_p=\Omega_p^{\star}$=1.77$\Omega$, with $g$=0.1$\Omega$. In Figs.~\ref{fig2} and 3, the strength of the laser pump and the FF intensity $|\alpha_1|^2$ do not change.
\begin{figure}
\begin{center}
\includegraphics[width=0.47\textwidth, trim=0cm 1.0cm 0.8cm 0cm,clip]{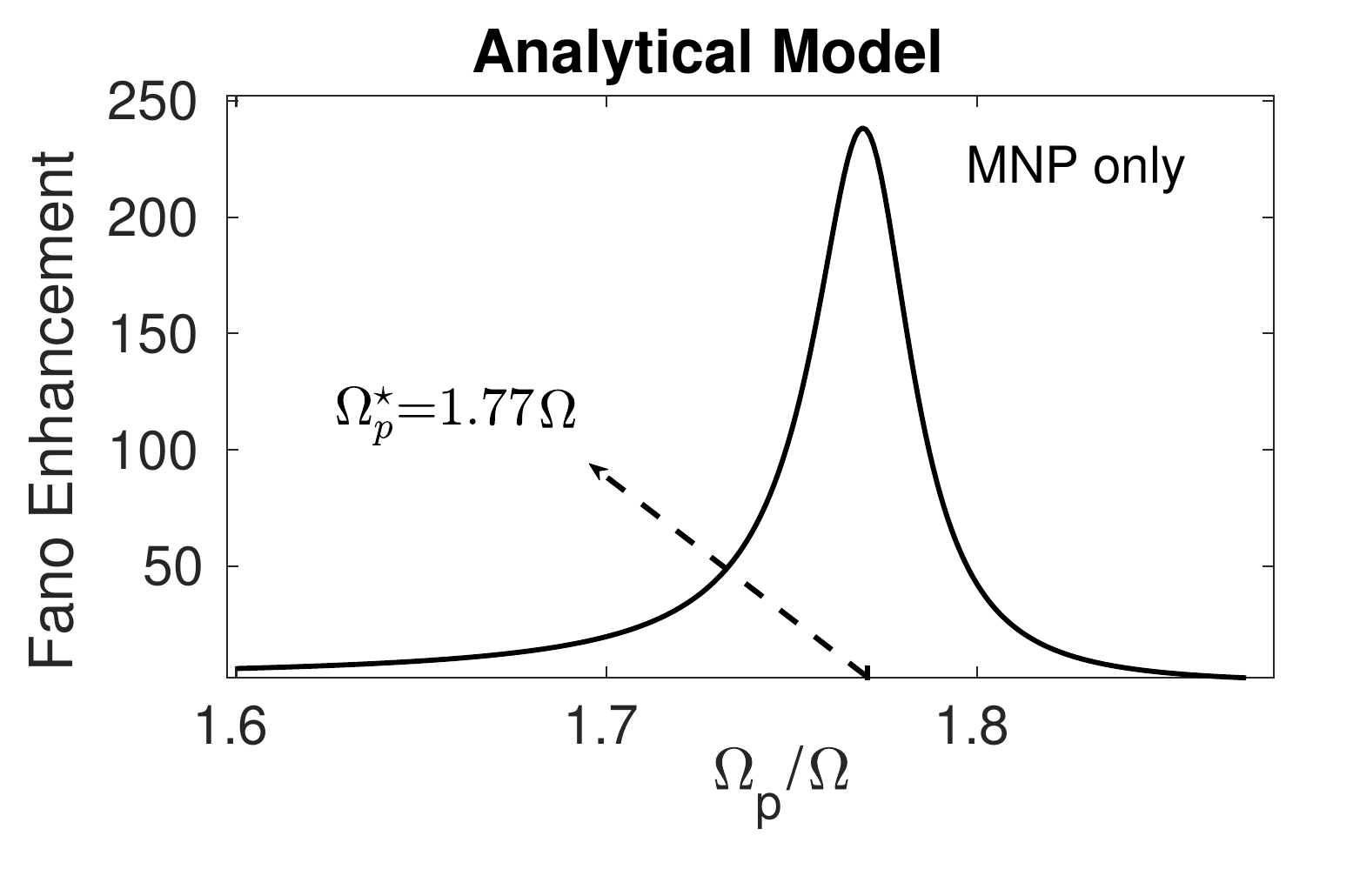}  
\caption{{\bf Fano enhancement can also be achieved using only MNPs.} We scan the plasmon resonance $\Omega_p$ of the MNP in order to find the particular value $\Omega_p^{\star}\simeq$1.77$\Omega$ where cancellations in the denominator of $\alpha_2$ takes place, see Eq.~(\ref{alpha2_MNP}).
}
\end{center}
\label{fig3}
\end{figure}
We confirm this phenomenon with the numerical solutions of the Maxwell's equations in the next section.

\subsection{Numerical solutions of Maxwell's equations}

In the previous section, we anticipated the presence of a 3 orders of magnitude Fano enhancement from our analytical model. In this section, we check the phenomenon with the numerical simulations of the Maxwell's equations, and we compare the two results. 

\begin{figure*}
\begin{center}
\includegraphics[width=150mm]{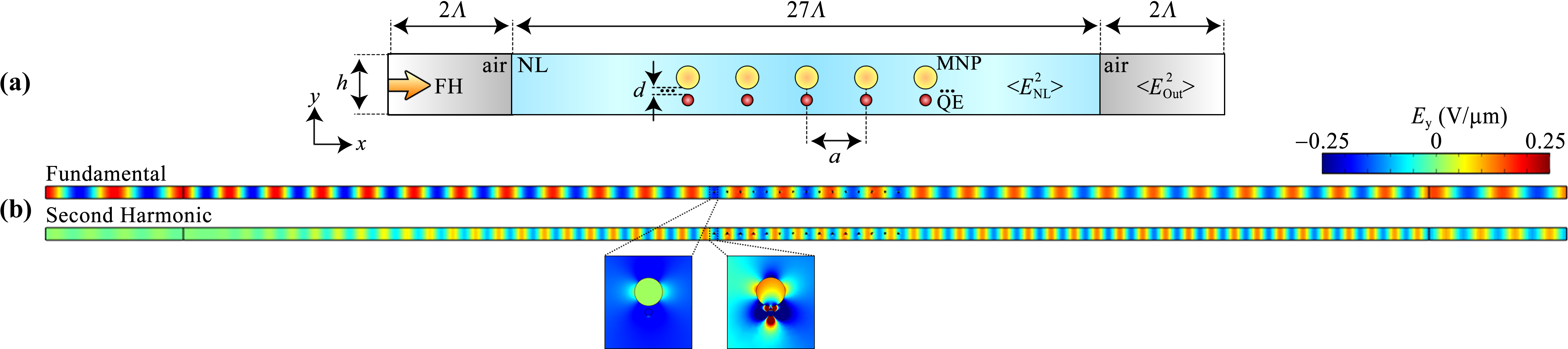}
\caption{{\bf A sketch of the computational domain employed in FEM simulations.} (a) A 27$\times$1064 nm long nonlinear (KDP) crystal is embedded with MNP-QE dimers of diameters 20 nm and 5 nm, respectively. The MNP and QE has a surface-to-surface distance of $d=$2 nm and the dimers are separated by $a$=200 nm in the x-direction. The E-field of the pump laser is chosen along the y-direction, thus the hot spot of the MNP is located at the position where the QE is placed. We use the experimental dielectric functions of the materials and a Lorentzian for the QE. (b) Close-up views of the FF and the SH field distributions. See the Supplementary Materials for further details.}
\label{fig4}
\end{center}
\end{figure*}

We solve the Maxwell's equations for the system presented in Fig.~\ref{fig4}a using finite element method~(FEM). We perform the calculations in COMSOL Multiphysics~\cite{comsol}. We consider a KDP nonlinear crystal for the SHG process. It has a refractive index of $n_{\scriptscriptstyle{\rm KDP}}$=1.51 and a second-order nonlinear coefficient of $d_2{\rm = 10^{-18}}$ ${\rm C/V^2}$. The length of the KDP crystal is chosen as 27$\times$1064 nm. We consider a width $h=$200 nm for the KDP in our simulations. The horizontal (the y-direction) boundaries are paired by Bloch-Floquet boundary conditions. 

The MNP-QE dimers embedded into the crystal have diameters of 20 nm and 5 nm, respectively. The distance between the MNP and QE is $d$=2 nm. The separation between the adjacent dimers is $a$=200 nm along the x-direction. We use the experimental dielectric functions for the KDP and the gold NP, and a Lorentzian dielectric function for the QE~\cite{wu2010quantum}. The linewidth of the QE is set as $\gamma_{\scriptscriptstyle{\rm QE}}$=5$\times {\rm 10^{11}}$ Hz. The QE has an oscillator strength of $f_{\rm osc}=0.1$ and the permittivity constant is set to $\epsilon_{\infty}$=1.

\begin{figure}
\centering
\includegraphics[width=0.47\textwidth]{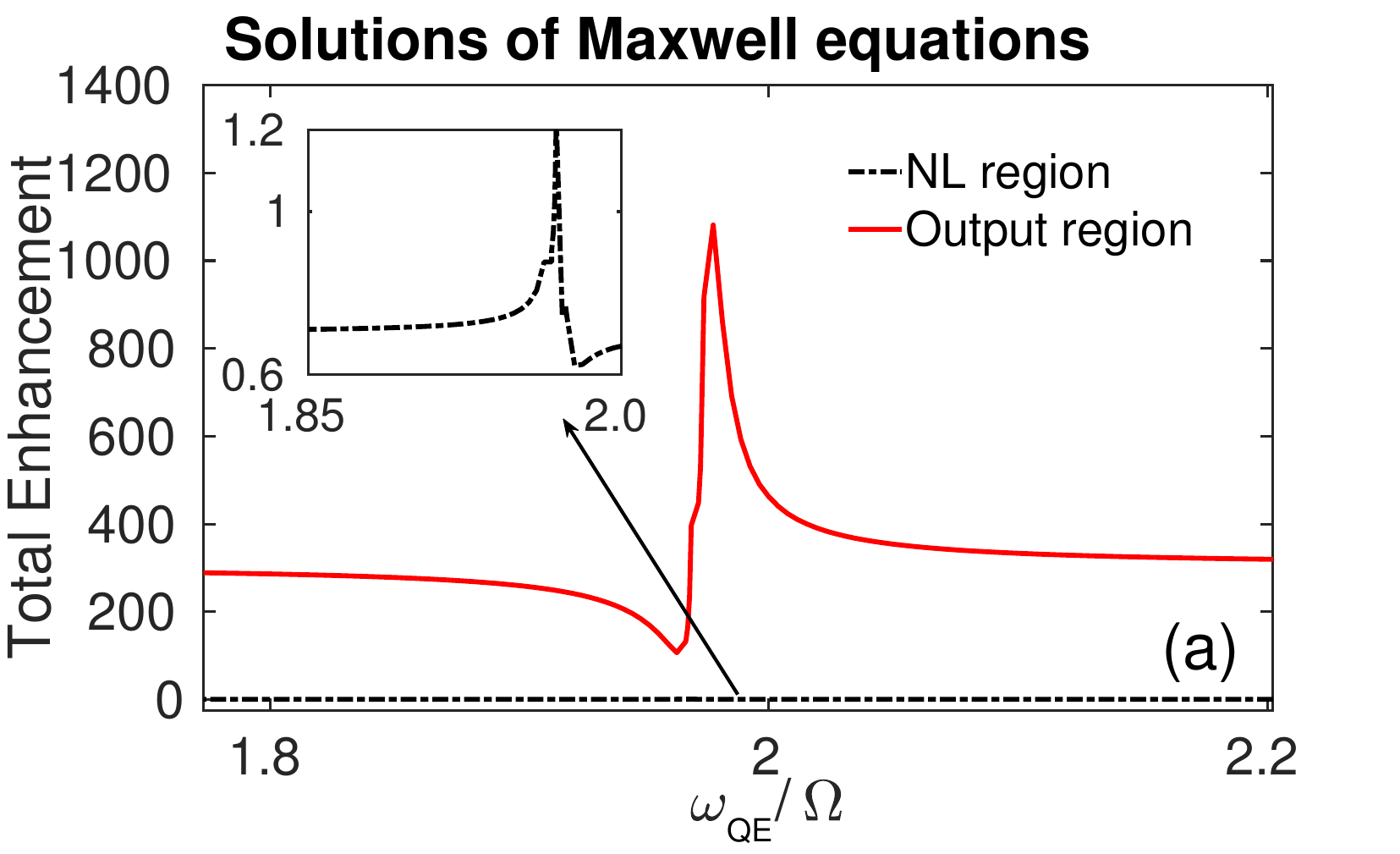}
\includegraphics[width=0.47\textwidth]{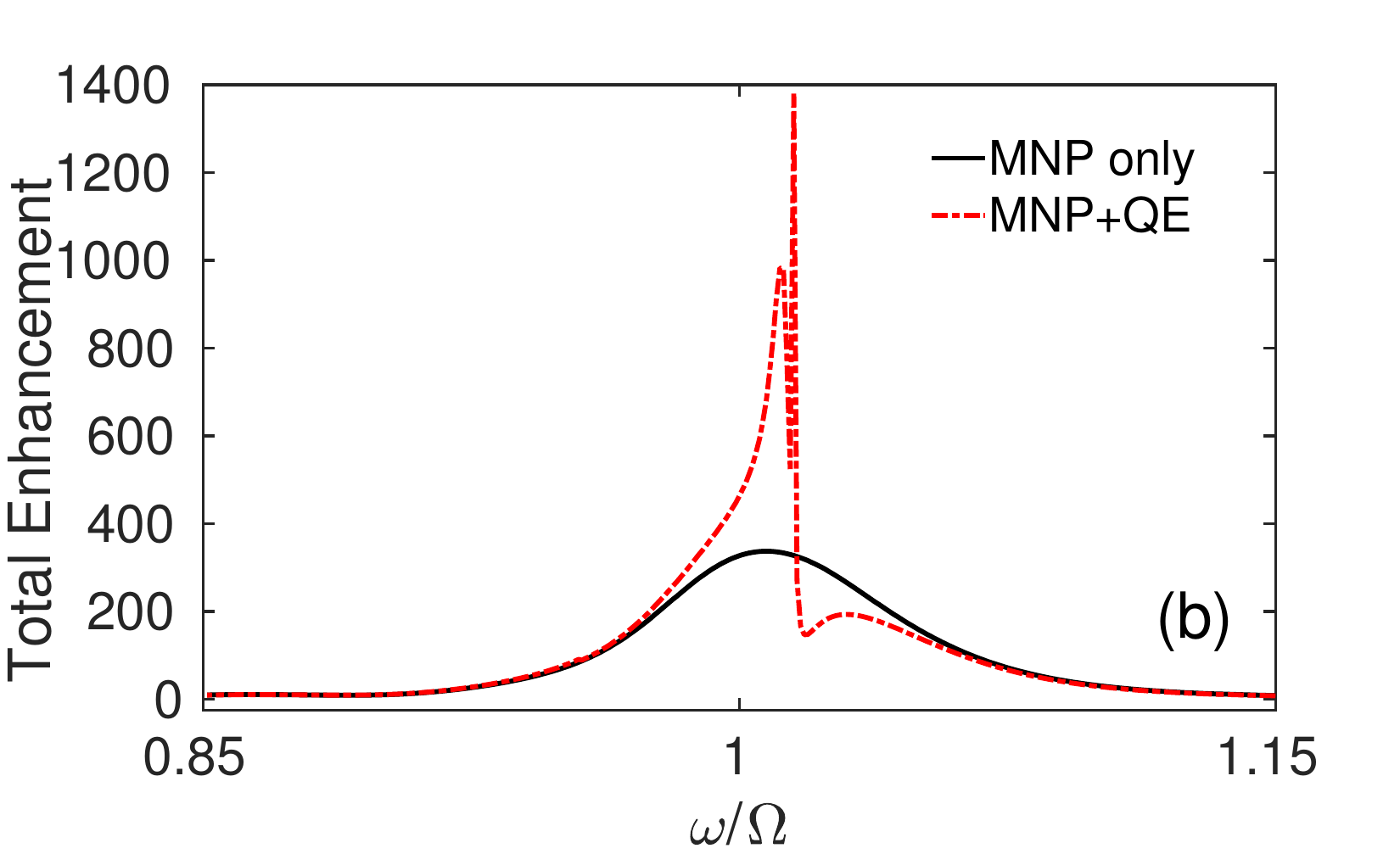}
\caption{ {\bf Numerical solutions of the Maxwell's equations.}  Total enhancement factors~(EFs) for the SH intensity, calculated at the crystal output, when MNP-QE dimers are placed into the crystal, see Fig.~\ref{fig4}. As the FF ($\omega$) intensity is not enhanced in the crystal, the plotted total EFs correspond to the Fano enhancement, see the text.  (a) Like in the analytical model, Fig.~\ref{fig2}a, we scan the QE level spacing $\omega_{\scriptscriptstyle{\rm QE}}$. A $\sim$1000 SH intensity enhancement takes place at $\omega_{\scriptscriptstyle{\rm QE}}$=$\omega_{\scriptscriptstyle{\rm QE}}^{\star}\simeq$1.98$\Omega$ which is similar with the analytical result in Fig.~\ref{fig2}a. (b) We fix the $\omega_{\scriptscriptstyle{\rm QE}}$=2$\Omega$, as in Fig.~\ref{fig2}b, and scan the pump frequency $\omega$. A $\sim$1400 enhancement factor appears (red dashed line) at $\omega=\omega^{\star}$=1.015$\Omega$ which is close to the one predicted by the analytical model in Fig.~\ref{fig2}b. Fano enhancement also takes place using  only MNPs (black line) as the analytical model predicts in Fig.~\ref{fig2}b (black line). The local fields are not enhanced.
}
\label{fig5}
\end{figure}

We pump the system with a laser of intensity $I_0$= 100 MW/${\rm m^2}$. The wavelength of the pump is fixed at $\lambda$=1064 nm ($\omega=2\pi c/\lambda$) in Fig.~\ref{fig5}a, but varied in Fig.~\ref{fig5}b. As in the analytical treatment, we scale frequencies by the frequency corresponding to the 1064 nm laser, i.e., $\Omega=2\pi c/\Lambda$ with $\Lambda$= 1064 nm. Hence, in Fig.~\ref{fig5}a the laser frequency is set to $\omega={\rm 1}\times\Omega$. Details about the computations can be found in the Supplementary Materials.

\paragraph*{Fano Enhancement}

We calculate the SH field at (i) the output, (ii) along the crystal body and (iii) at the hot spots between the MNP and the QE, see Fig.~\ref{fig4}a. Simulations produce interesting results.

In Fig.~\ref{fig5}, we plot the total SH enhancement at the ``output'' of the KDP crystal, i.e., at the air region on the right in Fig.~\ref{fig4}a. We compare the SH intensities with and without the presence of the MNP-QE dimers.  In Fig.~\ref{fig5}a, we fix the pump frequency to $\omega=\Omega$ and seek for the optimum value of the QE level spacing $\omega_{\scriptscriptstyle{\rm QE}}^{\star}$ at which we obtain an appreciable enhancement. At $\omega_{\scriptscriptstyle{\rm QE}}=\omega_{\scriptscriptstyle{\rm QE}}^{\star}\simeq$1.98$\Omega$, a $\sim$1000 times SHG enhancement at the KDP output can be observed. In contrary to such a total enhancement, the average SH field inside the crystal is suppressed to the 70\% of the one for the bare crystal~(without the MNP-QE dimers). The SH intensity in the crystal is enhanced only at the hot spot between the MNP and the QE. But this is only a 2 times enhancement and takes place only in a small region. Moreover, the FF $\omega$ intensity inside the crystal is 95\% of the one for the bare crystal.

Therefore, the $\sim$1000 times total enhancement at the crystal output cannot originate from the localization (field trapping) type enhancement employed in other studies~\cite{nie2018plasmonic,li2017giant,ENZnonlinearNatureReview2019,capretti2015comparative,luk2015enhanced,deng2020giant,vincentiLENZ2017}. Then, the observed SHG enhancement is the Fano enhancement that our analytical model predicts in Fig.~\ref{fig2}a. Furthermore, the Fano enhancement in the analytical model and the numerical simulations appears at the same QE level spacing $\omega_{\scriptscriptstyle{\rm QE}}=\omega_{\scriptscriptstyle{\rm QE}}^{\star}\simeq$1.98$\Omega$. We make an effort to use similar parameters in the analytical and numerical treatments. For instance, the resonances $\Omega_1\simeq$1.007$\Omega$ (FF mode) and $\Omega_2\simeq$2.013$\Omega$ (SH mode), which we use in the analytical model, are chosen by rough calculations for the KDP crystal depicted in Fig.~\ref{fig4}a. Our purpose in conducting the numerical simulations is only to demonstrate the presence of the Fano enhancement in the SHG process, in the absence of field trapping. 



We also compare the analytical and numerical results for a varying laser (pump) frequency $\omega$, at the fixed QE level spacing $\omega_{\scriptscriptstyle{\rm QE}}$=2$\Omega$. Fig.~\ref{fig5}b shows that a $\sim$1500 times total SHG enhancement appears at the ``output'' of the nonlinear crystal. This enhancement is attained at the pump frequency $\omega=\omega^{\star}$=1.015$\Omega$ ($\lambda^{\star}$=1048 nm). Please note that the enhancement peak at $\omega=\omega^{\star}$=1.01$\Omega$ (Fig.~\ref{fig2}b), in the analytical treatment, shows an excellent agreement with the numerical results. Additionally, both Fig.~\ref{fig5}b and Fig.~\ref{fig2}b  display a shouldered curve as a common aspect.

The average SH intensity in the crystal body drops to the 90\% of the one for the bare crystal. The SH intensity increases only 2 times at the small region, i.e., at the hot spot between the MNP and the QE. Moreover, the FF $\omega$ field is suppressed to the 56\% of the bare crystal. The localized field enhancement is not sufficient to be accounted for the SHG enhancement alone.  Thus, the $\sim$1500 times total enhancement at the output can only be explained by the unlocalized Fano enhancement factor as we suggest.

In our FEM simulations, the dimers are placed periodically ($a$=200 nm) along the x-direction. We also checked if any effect due to periodicity intervenes with the Fano enhancement profile in Fig.~\ref{fig5}. We performed control simulations also for different distances between the dimers other than $a$=200 nm. The enhancement factors remained almost unaltered.

\paragraph*{Fano enhancement using only MNPs}

We also carry out FEM simulations for a KDP crystal embedded only with MNPs. We confirm the Fano enhancement effect predicted by the analytical model in Eq.~(\ref{alpha2_MNP}). In Fig.~\ref{fig5}b, black line, we observe that the total SH intensity at the crystal output is enhanced about 300 times using only MNPs. This enhancement takes place in spite of the fact that both FF and SH fields inside the crystal do not intensify. That is, the 300 times total enhancement cannot be explained with the localization effects, thus it is chiefly due to the Fano enhancement effect. Comparing the black lines in Fig.~\ref{fig5}b and Fig.~\ref{fig2}b, one can observe that the Fano enhancement appears at similar locations. A comparison with Fig.~3 cannot be performed since the actual (experimental) dielectric function for the gold NP is used.

As a final remark, the resonance of the MNPs ---we use in all of the analytical and numerical simulations--- are set around the SH $2\omega$ field. That is, one already should not expect a localization effect in the fundamental ($\omega$) field. The localization in the SH field ---only at the hot spots--- is at most 2 times in all of the FEM simulations, thus cannot explain the 3 order enhancements at the crystal output. We remind that the average SH intensity inside the crystal even reduces compared to a bare crystal.

\section{Discussion}

We demonstrate the Fano control of unlocalized nonlinear processes which take place throughout an entire crystal body. MNP-QE dimers interact with the converted crystal field and introduce Fano resonances. The dimers can either be embedded into the crystal or they can be placed on the crystal surface. A 3 orders of magnitude Fano enhancement takes place for the appropriate choice of the QE's level spacing, $\omega_{\scriptscriptstyle{\rm QE}}$. We demonstrate the phenomenon both on a basic analytical model and via FEM-based numerical solutions of the Maxwell's equations.

The analytical model predicts the presence of such a Fano enhancement and the FEM simulations confirm the existence of the Fano enhancement. The analytical model yields a simple expression for the SH field amplitude, see Eq.~(\ref{alpha2}). The mechanism of the enhancement can be explained on this simple expression: the cancellations in the denominator originates the Fano enhancement. The expression also shows why the Fano enhancement multiplies the enhancements due to field trapping techniques~\cite{nie2018plasmonic,li2017giant,ENZnonlinearNatureReview2019,capretti2015comparative,luk2015enhanced,deng2020giant,vincentiLENZ2017}.

The FEM simulations demonstrate the Fano enhancement predicted by the analytical model. The crystal output field (in total) is enhanced by 3 orders, while the linear and nonlinear fields inside the crystal are not enhanced. Thus, the 3 orders of magnitude total enhancement is proven to originate from the Fano enhancement. Moreover, the frequency dependence of the Fano enhancement factors display a good match between the analytical and the FEM results.

The Fano enhancement scheme we develop is outstanding, because (i) it multiplies ($\sim10^3$) the enhancements achieved via field trapping techniques, (ii) an applied voltage can continuously tune the Fano resonances, and (iii) of its implementations with portable devices. Moreover, (iv) the extra (Fano) enhancement scheme works equally for other nonlinear processes, see the Supplementary Materials.

 The Fano enhancement scheme can be used together with any of the field trapping techniques. It multiplies the enhancements achieved by ENZ materials~\cite{ENZnonlinearNatureReview2019,capretti2015comparative,luk2015enhanced,deng2020giant,vincentiLENZ2017} or nanoparticle doping~\cite{nie2018plasmonic,li2017giant} by a factor of $\sim{}10^3$. ENZ compatibility of the method has particular importance for longer photonic devices. This is because, ENZ materials relieve the phase matching conditions substantially~\cite{kauranen2013phasematching}, thus they are unrivaled for the efficient operation of nonlinear devices longer than $\sim$10 $\mu$m~\cite{kauranen2013phasematching}.




The level spacing of a QE can be tuned via an applied voltage~\cite{schwarz2016electrically}, thus one can continuously switch between different Fano enhancement factors in Figs.~\ref{fig2}a and \ref{fig5}a. Fano enhancements appear at sharp resonances. Roughly, an 8-10 meV voltage tuning~\cite{QEtuning2005} corresponds to an arrangement of the Fano enhancement factor between 100-1000. 


 The demonstrated unlocalized Fano enhancement scheme is quite important also for portable device implementations. Considering that Fano enhancement multiplies the SH signal by a factor of $\sim$1000, the same SH intensity can be obtained using a $I_0/\sqrt{1000}\simeq I_0/33$ intensity laser, instead of an $I_0$ laser. Thus, battery life of such a portable laser can be extended significantly.

 In the paper, we only work on the Fano enhancement of the SHG process. The Fano enhancement scheme that we demonstrate, however, can also be used in other nonlinear processes. In the Supplementary Materials, we show that Fano enhancement schemes, similar to Eq.~(\ref{alpha2}), appear also for the third harmonic generation and the four-wave mixing processes. Similarly, Fano enhancement further multiplies the enhancements achieved by field trapping techniques in higher order nonlinearities.

\section*{Appendix}

In this section, we obtain the equations of motion~(EOM) for the dynamics of a nonlinear crystal coupled to a MNP-QE dimer.

The components of the hamiltonian for the coupled system can be expressed as follows. The occupation energies of the crystal modes ($\alpha_1$, $\alpha_2$), the plasmon mode~($\alpha_p$) and the excitation of the QE can be written as
\begin{equation}
\hat{\cal H}_0=\hbar \Omega_1 \hat{a}_1^\dagger \hat{a}_1+\hbar \Omega_2 \hat{a}_2^\dagger \hat{a}_2+\hbar \Omega_p \hat{a}_p^\dagger \hat{a}_p+\hbar \omega_{eg} |e\rangle \langle e|, \nonumber
\end{equation}
where $\hat{a}_{1,2}^\dagger$ ($\hat{a}_{1,2}$) creates (annihilates) a photon in the $\alpha_{1,2}$-modes of the nonlinear crystal. Similarly, $\hat{a}_p^\dagger$ ($\hat{a}_p$) creates (annihilates) a plasmon in the $\alpha_p$-mode of the MNP. The field profile of the MNP's plasmon mode displays a localization at its hot spot. $|e\rangle$ stands for the excited state and the operator $|e\rangle\langle e|$ determines the probability of the QE to be in the excited state. $\hat{a}_i^\dagger \hat{a}_i$, $i=1,2,p$, determines the number of photons (plasmons) in the modes.

The $\alpha_1$-mode of the crystal is pumped 
\begin{equation}
\hat{{\cal H}}_{\rm {\scriptscriptstyle{L}} }=i\hbar (\varepsilon_{\rm \scriptscriptstyle{L}}  \hat{a}_1^\dagger e^{-i\omega t} -\textit{h.c.})
\end{equation}
by a laser of frequency $\omega$. Thus, $\omega$ frequency photons are created~($\hat{a}_1^\dagger$) in the $\alpha_1$-mode. $\varepsilon_{\rm \scriptscriptstyle{L}}$ is proportional to the laser field amplitude. The ${\it h.c.}$ stands for the hermition conjugate. The crystal performs SHG: two $\omega$ photons in the $\alpha_1$-mode annihilates~($\hat{a}_1^2$) and a $2\omega$ photon is created in the~($\hat{a}_2^\dagger$) $\alpha_2$-mode
\begin{equation}
\hat{{\cal H}}_{\scriptscriptstyle \rm SH} = \hbar \chi_2 (\hat{a}_2^\dagger \hat{a}_1 \hat{a}_1 + {\it h.c.} ).
\end{equation}
The converted ($2\omega$) field couples with the plasmon mode of the MNP
\begin{equation}
\hat{\cal H}_{g}=\hbar g(\hat{a}_p^\dagger \hat{a}_2+ {\it h.c.}),
\end{equation}
where a photon in the $\alpha_2$-mode annihilates~($\hat{a}_2$) and creates a plasmon~($\hat{a}_p^\dagger$) on the MNP ($\alpha_p$-mode). The plasmon mode profile dispays a hot spot. The QE placed in at the hot spot interacts strongly with the plasmon field. A plasmon can annihilate~($\hat{a}_p$) and excite the QE to the upper level~($|e\rangle \langle g|$)
\begin{equation}
\hat{\cal H}_{\rm int}= \hbar f ( |e\rangle \langle g| \hat{a}_p  + {\it h.c.}).
\end{equation}
Thus, the total hamiltonian can be written as
\begin{equation}
{\hat{\cal H}}= \hat{\cal H}_0 + \hat{{\cal H}}_{\rm {\scriptscriptstyle{L}} } + \hat{{\cal H}}_{\scriptscriptstyle \rm SH} + \hat{\cal H}_{g} + \hat{\cal H}_{\rm int}.
\label{Htot}
\end{equation}

The set of EOMs for this system can be obtained using the Heisenberg EOM, e.g., $i\hbar \dot{a}=[\hat{a}_i,\hat{\cal H}]$ where $i=1,2,p$. For investigating only the field amplitudes,   the second-quantized operators can be replaced by their expectations, e.g., $\alpha_i=\langle \hat{a}_i\rangle$. We also replace $\hat{\rho}_{ij}=|i\rangle\langle j| \to \rho_{ij}$~\cite{Premaratne2017}. Including also the decay rates for the fields and the QE, the EOM can be obtained as
\begin{subequations}
\begin{align}
\dot{{\alpha}_1}&=-(i\Omega_1+\gamma_{1}) {\alpha}_1-i 2 \chi_2 {\alpha}_1^\ast {\alpha}_2+ \varepsilon_{\scriptscriptstyle \rm L} e^{-i\omega t}, \label{EOMa}\\
\dot{{\alpha}_2}&=-(i\Omega_2+\gamma_{2}) {\alpha}_2-i  \chi_2 {\alpha}_1^2 -i g {\alpha}_p,\label{EOMb} \\
\dot{{\alpha}}_p &=  -(i \Omega_p+\gamma_{p}){\alpha}_p-i g {\alpha}_2-if{\rho}_{ge}, \label{EOMc}\\
\dot{{\rho}}_{ge} &=  -(i \omega_{eg}+\gamma_{eg}){\rho}_{ge}+i f {\alpha}_p({\rho}_{ee}-{{\rho}}_{gg}), \label{EOMd}\\
\dot{{{\rho}}}_{ee} &= -\gamma_{ee} {{\rho}}_{ee}+i (f {\rho}_{ge} {\alpha}^\ast_p- \textit{c.c}),
\label{EOMe}
\end{align}
\end{subequations}
where $\gamma_i$ stands for the decay rates and $\gamma_{eg}=\gamma_{ee}/2$. In the text, we use $\gamma_{\rm \scriptscriptstyle QE} \equiv \gamma_{eg}$ for simplicity. 

The system is driven by a source oscillating as $\sim e^{-i\omega t}$. Investigating Eqs.~(\ref{EOMa})-(\ref{EOMe}), one can see that the solutions at the steady state oscillate as ${\alpha}_1(t)=\tilde{\alpha}_1 e^{-i\omega t}$, ${\alpha}_{2,p}(t)=\tilde{\alpha}_{2,p} e^{-i2\omega t}$ and ${\rho}_{ge}(t)= \tilde{\rho}_{ge} e^{-i2\omega t}$, where the terms with the tilde~($\tilde{\;}$) are constants which determine the steady state field amplitudes. We put these solutions into the EOM~(\ref{EOMa})-(\ref{EOMe}) and obtain the equations 
\begin{subequations}
\begin{align}
0&=-[i(\Omega_1-\omega)+\gamma_1 ] \tilde{\alpha}_1-i 2 \chi_2 \tilde{\alpha}_1^\ast \tilde{\alpha}_2+ \varepsilon_{\scriptscriptstyle \rm L}, \label{stEOMa}\\
0&=-[i(\Omega_2-2\omega)+\gamma_2 ] \tilde{\alpha}_2-i  \chi_2 \tilde{\alpha}_1^2 -i g \tilde{\alpha}_p,\label{stEOMb} \\
0 &=-[i( \Omega_p-2\omega)+\gamma_{p}]\tilde{\alpha}_p-i g \tilde{\alpha}_2-if \tilde{\rho}_{ge}, \label{stEOMc}\\
0&=  -[i (\omega_{eg}-2\omega)+\gamma_{eg}]\tilde{\rho}_{ge}+i f \tilde{\alpha}_p({\rho}_{ee}-{{\rho}}_{gg}), \label{stEOMd}\\
0 &= -\gamma_{ee} {{\rho}}_{ee}+i (f \tilde{\rho}_{ge} {\alpha}^\ast_p- f^*\tilde{\rho}_{ge}^* {\alpha}_p)
\label{stEOMe}
\end{align}
\end{subequations}
for the steady state values.  The equations for the steady state amplitudes are not exactly solvable. However, one can still express the SH field amplitude $\tilde{\alpha}_2$ as in Eq.~(\ref{alpha2}), using Eqs.~(\ref{stEOMb})-(\ref{stEOMd}). This solution, i.e., Eq.~(\ref{alpha2}), provides us an invaluable insight for the Fano enhancement and its relation to the enhancements obtained via field trapping techniques.  In the text, in Eqs.~(\ref{alpha2}) and (\ref{alpha2_MNP}), we drop the tilde symbols ($\tilde{\;}$) for a simple presentation.


\nocite{zernike1964refractive}
\nocite{johnson1972optical}
\nocite{thomas2018plexcitons}
\nocite{leistikow2009size}

\begin{acknowledgements}
MG and MET are supported by TUBITAK-1001 under grant no 117F118. AB and MET are supported by TUBITAK-1001 under grant no 119F101. MET and AC acknowledge support from Turkish Academy of Sciences~(TUBA) Outstanding Young Researchers Awarding Programme (GEBIP).
\end{acknowledgements}

\bibliography{bibliography}

\end{document}